\documentclass[a4paper]{article}
\usepackage[utf8]{inputenc}
\usepackage{graphics}
\usepackage{amsmath}
\usepackage{mathtools}
\usepackage{authblk}
\usepackage[a4paper,left=0.65in,top=0.65in]{geometry}
\usepackage{hyperref}
\usepackage{xcolor}
\newcommand{\ours}{{\sc LoMeX}}
\newtheorem{theorem}{Theorem}
\newtheorem{problem}[theorem]{Problem}
\newtheorem{definition}[theorem]{Definition}

\begin{document}

\title{Extraction of long \emph{k}-mers using spaced seeds}

\author{Miika Leinonen and Leena Salmela}

\affil{Department of Computer Science,\\Helsinki Institute for Information Technology HIIT,\\ University of Helsinki\\\{miika.leinonen, leena.salmela\}@helsinki.fi}

\date{}

\maketitle

\begin{abstract}

The extraction of $k$-mers from sequencing reads is an important task in many bioinformatics applications, such as all DNA sequence analysis methods based on de Bruijn graphs. These methods tend to be more accurate when the used $k$-mers are unique in the analyzed DNA, and thus the use of longer $k$-mers is preferred. When the read lengths of short read sequencing technologies increase, the error rate will become the determining factor for the largest possible value of $k$. Here we propose \ours\ which uses spaced seeds to extract long $k$-mers accurately even in the presence of sequencing errors. Our experiments show that \ours\ can extract long $k$-mers from current Illumina reads with a higher recall than a standard  $k$-mer counting tool. Furthermore, our experiments on simulated data show that when the read length further increases, the performance of standard $k$-mer counters declines, whereas \ours\ still extracts long $k$-mers successfully.
\end{abstract}

\section{Introduction}

Counting and extracting $k$-mers, i.e.\ subsequences of length $k$, from sequencing reads is a frequently used technique in bioinformatics applications and many tools have been developed to solve the task \cite{kmercountmethods18}. A $k$-mer counter needs to enumerate all different subsequences of length $k$ that occur in the sequencing reads and report the frequency of each such $k$-mer.

Counting $k$-mers has several applications in bioinformatics. In the overlap-layout-consensus approach to genome assembly, $k$-mers can be used to identify candidate pairs of overlapping reads by finding reads that share a substantial amount of $k$-mers \cite{cabog, arachne}. These candidate read pairs are then further verified for actual overlaps by aligning them. In the de Bruijn graph-based approaches~\cite{zerbino08velvet,Simpson09abyss}, $k$-mer counting is the first step as it identifies the distinct $k$-mers occurring in the reads that will become the edges of the de Bruijn graph, whereas the $k-1$-mers will form the nodes.

Correction of sequencing errors in reads is another application where $k$-mer counting plays an important role. The correction procedure may be entirely based on the $k$-mer spectrum of the reads \cite{quake,musket,hammer} or $k$-mers can be used to filter reads for multiple alignments \cite{coral}. Other approaches rely on de Bruijn graphs which are built on $k$-mer sets \cite{salmela14lordec}. Other applications of $k$-mer counting include metagenomic classification \cite{kraken}, repeat classification \cite{rap,forrepeats}, and SNV calling directly from read data \cite{discosnp,fastgt}.

In many applications, it is important to be able to count long $k$-mers because the longer the $k$-mers are, the more likely they are unique in the genome. For example, a de Bruijn graph will be simpler with fewer branches when the $k$-mers are longer. On the other hand, if the $k$-mers are short, the de Bruijn graph will be more complex having multiple branching paths, which makes it difficult to infer which long sequences occur in the genome with high confidence. However, if $k$ is too big, the unique $k$-mer abundance starts to drop and the graph becomes too fragmented. The optimal choice of $k$ has been studied \cite{kmersize13}, but it is hard to estimate the best choice of $k$. Nevertheless, the usage of long $k$-mers can be beneficial in many bioinformatics applications. 

The accurate and short read sequencing technologies, such as Illumina, nowadays reach read lengths of 300 bp, which should allow the use of longer $k$-mers. However, although the error rate of these technologies is low, the sequencing errors will become a limiting factor for determining the largest possible value of $k$ in the standard $k$-mer counting methods when the read lengths further increase. These methods involve only counting how many times each $k$-mer appears in the data. Usually, a $k$-mer should occur at least twice for it to be counted as a real $k$-mer existing within the data. With short enough $k$-mers one expects to find enough error-free occurrences of the $k$-mers but the likelihood of finding at least two error-free occurrences of a $k$-mer decreases as $k$ increases. A higher coverage increases the likelihood of finding at least two error-free occurrences of a $k$-mer but producing high coverage data sets is more costly and might be infeasible if the required coverage is very high. 

Development of $k$-mer counting methods has largely concentrated on time and memory efficiency. Much less attention has been given to improving the quality, i.e. getting long and accurate $k$-mers. Here we propose \ours\ to extract long $k$-mers with high precision and recall. \ours\ uses \textit{spaced seeds}, which are patterns of length $k$, where $k$ is the length of the searched $k$-mers. Some of the $k$ positions in the pattern are fixed, while the others are gap positions, which are also known as "don't care" positions. When we search for a $k$-mer with this pattern, only the characters in the fixed positions are considered. If two $k$-mers have the same characters in their respective fixed positions, they are counted as the same \textit{spaced} $k$\textit{-mer}. Afterward, the consensus of the $k$-mers matching to the same spaced $k$-mer are used to fill in the "don't care" positions, resulting in a long \textit{consensus $k$-mer}.

We compared \ours\ to DSK~\cite{dsk13} which is a standard $k$-mer extraction tool. Our results show that on current Illumina data \ours\ typically has higher recall than DSK with a small drop in precision. Extracting $k$-mers with a high recall is important in downstream applications. A missing $k$-mer often cannot be recovered but it is possible to later discard erroneous $k$-mer information, for example by examining the tip and bubble structures of a de Bruijn graph.
Furthermore, our experiments on simulated data show that when read lengths further increase, the error rate will become the limiting factor for choosing a large $k$ in standard $k$-mer extraction, whereas \ours\ can still extract long $k$-mers successfully.

\ours\ is freely available at https://github.com/Denopia/LoMeX

\section{Related work}

\subsection{$k$-mer counting}

Many strategies have been developed to count the $k$-mers present in a set of sequencing reads. For example KMC3~\cite{kmc3}, Turtle~\cite{turtle}, and GenomeTester4~\cite{GenomeTester4} use sorting to count $k$-mers. In this approach, all $k$-mers are extracted from the reads. The $k$-mers are then sorted and from the sorted list of $k$-mers, it is easy to count how many times each $k$-mer occurs.

An alternative method is to use a data structure to store the $k$-mers and their counts. Jellyfish~\cite{jellyfish} implements a lock-free hash table using a compare-and-swap assembly instruction to store the $k$-mers as keys and the counts as values. The lock-free data structure enables fast, parallel $k$-mer counting.

Other tools have employed approximate membership query (AMQ) data structures for more efficient $k$-mer counting. BFCounter~\cite{bfcounter} uses Bloom filters to filter out singleton $k$-mers and stores the non-singleton $k$-mers and their counts in a hash table. To account for the false-positives of the Bloom filter, it reiterates over the sequence reads to correct the wrong counts in the hash table. Squeakr~\cite{squeakr17} uses counting quotient filters (CQF) to store the $k$-mer counts. Squeakr supports both exact and approximate $k$-mer counting.

Other data structures used by $k$-mer counters include enhanced suffix trees used by Tallymer~\cite{tallymer} and burst tries used by KCMBT~\cite{kcmbt}.

MSPKmerCounter~\cite{mspkmercounter}, KMC3~\cite{kmc3}, and Gerbil~\cite{gerbil} use minimum string partitioning to further reduce memory usage. They partition the input strings into multiple disjoint partitions based on minimizers and store several consecutive $k$-mers sharing a minimizer as a single super $k$-mer. The disjoint partitions can then be processed independently to get the actual $k$-mer counts.

Many $k$-mer counters use a disk-based implementation to save memory costs. For example, DSK~\cite{dsk13} first calculates the number of $k$-mer partitions it will need. The $k$-mers are then distributed to the partitions based on their hash values and an iteration number. The actual counting happens by loading one partition to memory at a time and counting the $k$-mers assigned to that partition. 

Also, GPU computation has been used to speed up $k$-mer counting~\cite{gerbil}. We refer the reader to \cite{kmercountmethods18} for a more detailed review of the various $k$-mer counting methods and their benchmarking.

\subsection{Spaced seeds}

Determining if two sequences are similar is a central question in biology. Initially, such problems were solved by pairwise alignment of the two sequences but the quadratic dynamic programming algorithms for pairwise alignment soon became too costly when the number of sequences increased as the number of pairwise comparisons also grows quadratically. The introduction of seeds presented a solution to this problem. The main idea is that similar sequences share identical regions and thus identical seeds can be found in these areas.

First programs for homology search, such as BLAST \cite{blast}, used matches of $k$-mers as seeds and then extended the seed matches to longer alignments. Spaced seeds \cite{patternhunter,buka03,ilil07,goodseeds,keli04} extend this concept: a spaced seed of length $k$ has a set of predefined positions that are required to match and the rest of the positions are so-called "don't care" positions that match any nucleotide. PatternHunter \cite{patternhunter} proposed to optimize the predefined positions required to match and obtained a significantly better sensitivity than BLAST \cite{blast}. Furthermore, Buhler et al.~\cite{buke05}, Ma et al.~\cite{patternhunter} and Brejová et al.~\cite{brbr03} noticed that using several spaced seeds further increased the sensitivity. In practice, spaced seeds have been shown to have high sensitivity and specificity for homology search even when the spaced seeds are not optimized \cite{patternhunter,patternhunter2}.

Spaced seeds are effective in finding similar sequences when the sequences mainly differ by mismatches. However, as the frequency of insertions and deletions increases, the length and the number of common spaced seeds found in similar sequences decreases, and the spaced seeds are no longer effective for identifying similar sequences which has been seen for example in long read alignment \cite{blasr}.

\section{Definitions}
\label{sec:definitions}

We start with a formal definition of the $k$-mer extraction problem. We then extend it first to spaced $k$-mer extraction and finally to consensus $k$-mer extraction. A {\em $k$-mer} is a sequence of $k$ characters. A {\em canonical $k$-mer} is a $k$-mer that is lexicographically smaller than its reverse complement. Canonical $k$-mers are used in $k$-mer counting because the reads can originate from either strand of the DNA molecule and we only want to count a $k$-mer once regardless of its orientation. Suppose we have 11-mer $m$ = ACTCATAATCA. Its reverse complement is $m'$=TGATTATGAGT, which is lexicographically bigger than $m$. Thus $m$ is a canonical $k$-mer, whereas $m'$ is not canonical. 

\begin{problem}[$k$-mer extraction] Given a set of reads $R$ and a threshold $S$, find all canonical $k$-mers that occur at least $S$ times in the reads and their reverse complements.
\end{problem}

First, we will extend the notion of a $k$-mer to a spaced $k$-mer. A {\em spaced seed pattern} is a string of zeros and ones where ones correspond to fixed characters, and zeros correspond to "don't care" characters. For example, spaced seed pattern 10010101001 could be used for searching 11-mers. We are now ready to define spaced $k$-mers.

\begin{definition}[Spaced $k$-mer]
A spaced $k$-mer $g$ adhering to a spaced seed pattern $p$ is a string of characters from \{A,C,G,T,*\} such that if $p[i] = 0$ then $g[i] = *$, and if $p[i] = 1$, then $g[i] \in \{A,C,G,T\}$.
\end{definition}

Because spaced seed patterns can be very long with long runs of zeros and ones, we will alternatively represent spaced seed patterns as a sequence of integers, where the numbers at odd positions indicate the number of consecutive fixed characters, and the numbers at even positions indicate the number of consecutive "don't care" characters. Thus the spaced seed pattern 10010101001 would be represented as 1-2-1-1-1-1-1-2-1.

As an example, let us assume we have the previous spaced seed pattern $p=10010101001$ and an 11-mer $m$ = ACTCATAATCA. Using pattern $p$ on this 11-mer $m$ yields spaced $k$-mer $g$ = A**C*T*A**A. If the spaced seed pattern is known, the "don't care" characters do not need to be included, and the spaced $k$-mer can simply be written as $g$ = ACTAA.

When the spaced seed pattern is palindromic, we can define a canonical spaced $k$-mer similar to canonical $k$-mers. A spaced $k$-mer is {\em canonical} if it is lexicographically smaller than its reverse complement. 

If the spaced seed pattern is not palindromic, the forward and reverse complementary spaced $k$-mers will contain characters from different positions. Thus the canonicality of a spaced $k$-mer would depend also on the characters in the "don't care" positions. For this reason, we will only consider palindromic spaced seed patterns from now on.

With this definition of canonical spaced $k$-mers, we can now define the spaced $k$-mer extraction problem:

\begin{problem}[Spaced $k$-mer extraction]
Given a set of reads $R$, a threshold $S$, and a spaced seed pattern $p$, find all canonical spaced $k$-mers that adhere to the pattern $p$ and occur at least $S$ times in the reads and their reverse complements.
\end{problem}

Each spaced $k$-mer reported by the solution to the spaced $k$-mer extraction problem has a set of occurrences in the reads. These occurrences determine which characters are solid at each position.  

\begin{definition}[Solid characters]
Let $Q$ be the set of occurrences of a spaced k-mer $g$ in a set of reads $R$ and let $c$ be the solidity threshold. Character $n_i \in \{A,C,G,T\}$ is a solid character at position $i$ of spaced $k$-mer $g$, if $n_i$ appears at least $c$ times at position $i$ in $Q$. 
\end{definition}

Each position of the spaced $k$-mer is classified as unambiguous, ambiguous or undecided. The number of solid characters determines how a position is classified. A position is unambiguous if it has only one solid character, and if there is more than one solid character, the position is classified as ambiguous. If there are no solid characters, the position is undecided. We note that all fixed positions of a spaced $k$-mer are unambiguous, whereas "don't care" positions can be in any of the three categories.

\begin{definition}[Unambiguous, ambiguous and undecided positions]
Position $i$ in spaced $k$-mer $g$ is unambiguous if it has exactly one solid character. If there are no solid characters, the position is undecided. Otherwise, the position is ambiguous.
\end{definition}

\begin{definition}[Consensus $k$-mer]
Consensus $k$-mer $d$ is a $k$-mer which corresponds to spaced $k$-mer $g$ such that $d[i]$ is a solid character of $g$ at position $i$, and there are at least two occurrences of the spaced $k$-mer where the ambiguous positions match exactly.
\end{definition}

Finally we are ready to define the consensus $k$-mer extraction problem:

\begin{problem}[Consensus $k$-mer extraction]
Find all consensus $k$-mers corresponding to all spaced $k$-mers found in a set of reads $R$.
\end{problem}

The reliability of $k$-mers is measured by {\em $k$-mer counts}, i.e. the number of occurrences a $k$-mer has in the read set. Here we extend the notion of $k$-mer counts to consensus $k$-mer {\em support counts}. Each occurrence of the spaced $k$-mer where the ambiguous positions match a consensus $k$-mer exactly contributes to the support count of that consensus $k$-mer. Other occurrences of the spaced $k$-mer are evenly distributed among all consensus $k$-mers of that spaced $k$-mer. 

\begin{definition}[Consensus $k$-mer support counts]
Let $Q$ be the set of occurrences of a spaced $k$-mer $g$ and $D$ the set of consensus $k$-mers of $g$. Given a consensus $k$-mer $d \in D$, we denote by $Q_d$ the set of occurrences where the ambiguous positions match $d$. The support count for a consensus $k$-mer $d$ is $|Q_d| + |(Q\setminus \bigcup_{d'\in D} Q_{d'})|/|D|$.
\end{definition}

\section{Methods}

\begin{figure*}[t]
    \centering
    \includegraphics[width=1.0\textwidth]{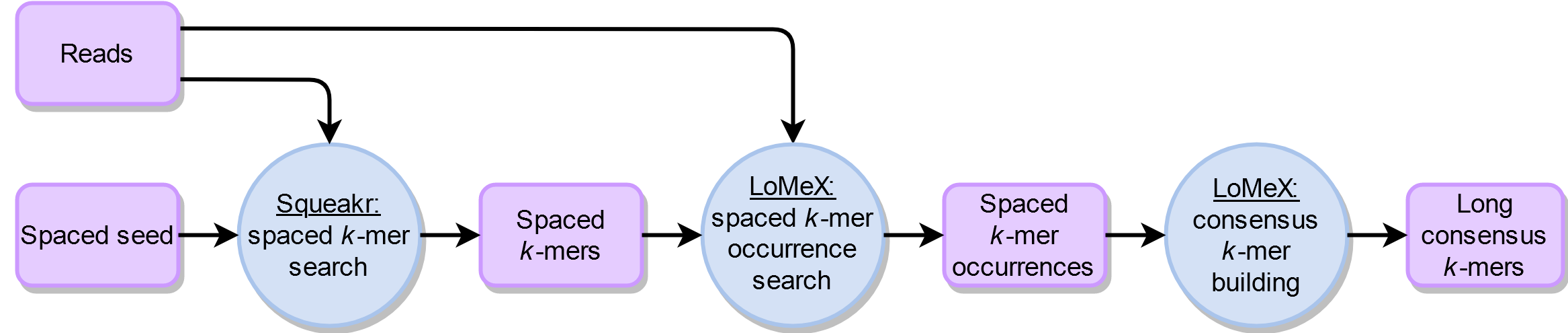}
    \caption{Long $k$-mer extraction steps. First, the spaced $k$-mers are extracted from the reads according to the chosen spaced seed pattern using a modified version of Squeakr $k$-mer extraction program. Next, the reads are scanned once more while the extracted spaced $k$-mers are kept in memory. For each spaced $k$-mer reported by Squeakr, their corresponding regular $k$-mers are stored on disk. Finally, the long $k$-mers are built based on the consensus of the regular $k$-mers.}
    \label{flowchart-img}
\end{figure*}

\ours\ pipeline can be divided into three steps. The first step solves the spaced $k$-mer extraction problem, the second step gathers $k$-mers corresponding to the extracted spaced $k$-mers, and the third step solves the consensus $k$-mer extraction problem for each extracted spaced $k$-mer. In the first step, a given spaced seed pattern is used to extract spaced $k$-mers in the reads. In the second step, the reads are scanned to find all occurrences of the extracted spaced $k$-mers. The core idea of this step is to group similar $k$-mers in the reads based on the chosen spaced seed pattern. In other words, $k$-mers that yield the same spaced $k$-mer belong in the same group. In the third step, the long consensus $k$-mers are built with the help of the grouped $k$-mers. More specifically, we use the consensus of the grouped $k$-mers to fill the "don't care" positions in the spaced $k$-mers to produce consensus $k$-mers, which are then reported as the output. Figure \ref{flowchart-img} depicts all the steps of this process, which are explained in more detail in the following sections.

\subsection{Spaced $k$-mer extraction}

In the first step, \ours\ extracts all canonical spaced $k$-mers that adhere to a given spaced seed pattern. As noted before, we are only considering palindromic spaced seed patterns. Furthermore, we use an odd number of fixed characters so that the spaced $k$-mers have an odd length. This ensures that a sequence and its reverse complement cannot be the same, so only one of them can be the canonical spaced $k$-mer.

To find all occurrences of the spaced $k$-mers in the input reads, we use a modified version of an existing $k$-mer counting system called Squeakr \cite{squeakr17}. Squeakr solves the $k$-mer extraction problem and thus, given a set of reads, threshold $S$, and the length $k$, it outputs all the $k$-mers that appear in the reads at least $S$ times. A $k$-mer is required to appear more than once to get rid of some of the spurious $k$-mers that arise due to the sequencing errors in the reads. By default, Squeakr reports $k$-mers that appear at least $S=2$ times.

We modified Squeakr so that instead of giving it the $k$-mer length, it takes a spaced seed pattern as input, which Squeakr then uses to report the corresponding spaced $k$-mers. Squeakr exact $k$-mer counting implementation supported only $k$-mers up to length 32. 
Because of this and the fact that we want to use odd length spaced $k$-mers, we are limited to using 31 fixed characters at most. There exists other $k$-mer counting tools that could work with longer $k$-mers, but Squeakr was the easiest one for us to modify for our needs, so we decided to use it over the other programs. 

\subsection{Long $k$-mer construction}

After Squeakr output has been obtained, the reported spaced $k$-mers are given to \ours\ which solves the consensus $k$-mer extraction problem. First, for each reported spaced $k$-mer, \ours\ finds its occurrences in the input reads. For each regular $k$-mer in the reads, \ours\ checks if the matching spaced $k$-mer appears in the Squeakr output. If this is the case, the regular $k$-mer is stored in memory associated with its spaced $k$-mer. Essentially, the $k$-mers in the read set are split into groups according to their corresponding spaced $k$-mers. 

Once all the spaced $k$-mers have been linked to their occurrences, i.e. corresponding regular $k$-mers, \ours\ fills the "don't care" (gap) positions of the spaced $k$-mers. \ours\ looks at the gap position characters of the regular $k$-mers and decides the characters for each gap position using the consensus of the regular $k$-mers. The number of regular $k$-mers stored for each spaced $k$-mer affects the quality of the constructed consensus $k$-mers. By default, there are at least two such $k$-mers, because Squeakr would not have reported the spaced $k$-mer otherwise. 

The consensus $k$-mers are constructed by filling the gap positions of the spaced $k$-mers with solid characters. For every spaced $k$-mer, the occurrences of the four possible characters (A, C, G, and T) in the regular $k$-mers are counted for each position. Then, the character counts are used to determine which characters are considered solid at each position. A character is solid if it appears at least $c$ times, where $c$ is the minimum character count threshold. To make \ours\ work on different sized inputs, we take the number of regular $k$-mers into account when deciding a suitable value for $c$. \ours\ requires that $c$ is at least ten percent of the number of all characters at the position, i.e. the character appears at least in ten percent of the regular $k$-mers at the specified position. Additionally, \ours\ has a hard minimum threshold $N$ for the number of required occurrences, set to two. The minimum character count threshold for spaced $k$-mer $g$ is defined in \ours\ as $c_g = \max(N, p \cdot |Q_g|)$, where $N=2$ is the absolute minimum number of required character occurrences, $p=0.1$ is the required proportion with respect to the number of regular $k$-mers, and $Q_g$ is the set of regular $k$-mers corresponding to $g$. The user can set different values for the parameters $N$ and $p$. The effect of $N$ on the accuracy of \ours\ is explored more in Section~\ref{experiments}.

\begin{figure*}[t]
    \centering
    \includegraphics[width=0.8\textwidth]{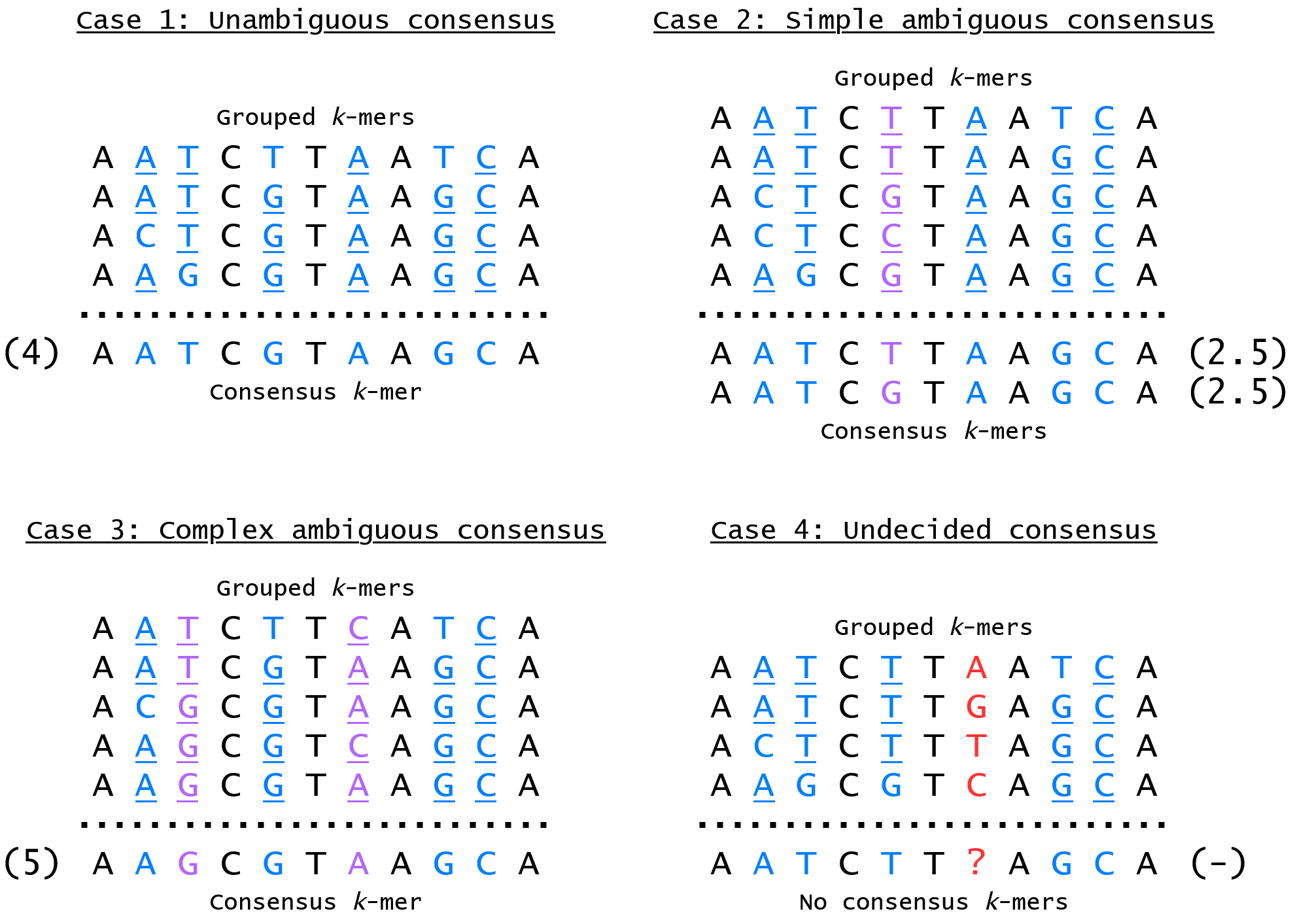}
    \caption{Four different gap-filling cases. In this example, we are filling the gaps of spaced $k$-mer A**C*T*A**A. Fixed characters are colored black. Gap characters are blue if only one character appears at least twice in its column. Gap characters are colored purple if there is more than one character which appears at least twice in its column. If none of the gap characters appears twice in its column, all the characters are colored red. Gap character is underlined if it appears more than once in its column. The numbers in brackets next to the consensus $k$-mers are their support counts.}
    \label{consensus-img}
\end{figure*}

Spaced $k$-mer "don't care" positions are only filled with solid characters. If only one of the characters has a high enough count to be solid, the position is unambiguous, and \ours\ simply uses that one to fill in the gap. If there are multiple potential character candidates at a position, it becomes ambiguous, making the whole consensus $k$-mer building case ambiguous. If there is a position with no solid characters, the consensus $k$-mer building is left undecided. \ours\ divides the consensus cases into four categories that are handled separately; unambiguous consensus, simple ambiguous consensus, complex ambiguous consensus, and undecided consensus. Examples of these cases are illustrated in Figure \ref{consensus-img}.

\begin{enumerate}
    \item \textbf{Unambiguous consensus.} Every position is unambiguous, i.e. there is only one character with a count greater than the required threshold $c$. The only valid characters are used to fill the gap positions, resulting in a single consensus $k$-mer.
    \item \textbf{Simple ambiguous consensus.} Only one position is ambiguous, i.e. there are multiple character candidates for it. All positions with only one valid character are filled as they were in the previous case, and the single ambiguous position is filled with all the valid character choices. At most, this results in four different consensus $k$-mers.
    \item \textbf{Complex ambiguous consensus.} There are multiple ambiguous positions. The consensus $k$-mers are solved by first filling in the unambiguous gap positions. Then \ours\ looks for regular $k$-mers that have the same characters at the ambiguous gap positions. If \ours\ finds at least two $k$-mers that share the same characters at the ambiguous positions, those characters are used to fill the remaining gaps to produce a consensus $k$-mer. This case gives us at most $\lfloor \frac{|Q_g|}{2} \rfloor$ consensus $k$-mers, where $Q_g$ is the set of regular $k$-mers associated with spaced $k$-mer g.
    \item \textbf{Undecided consensus.} There is at least one position where no character is solid. This can happen when the number of regular $k$-mers corresponding to a spaced $k$-mer is very low. This results in zero reported consensus $k$-mers.
\end{enumerate}{}

 The gaps in the spaced $k$-mers are filled with the consensus characters to produce the long consensus $k$-mers, which \ours\ then reports as its output. Even though we discarded the non-canonical spaced $k$-mers, this does not mean that the reported consensus $k$-mers are necessarily canonical. However, \ours\ only reports $k$-mers which correspond to canonical spaced $k$-mers and thus only a $k$-mer or its reverse complement is reported but never both. Therefore it is easy to transform the output of \ours\ so that only canonical $k$-mers are reported. One simply needs to check if a reported consensus $k$-mer is canonical and if it is not, report its reverse complementary sequence instead.

As their name implies, $k$-mer counters produce the extracted $k$-mers and how many times they appear in the input reads. We implemented a similar feature for \ours\ but instead of exact $k$-mer counts, \ours\ reports the support counts of the consensus $k$-mers as defined in Section \ref{sec:definitions}. The support counts tell how strongly a produced $k$-mer is supported by the spaced $k$-mer occurrences in the input data.

Because there are four different types of consensus $k$-mers, we must also define how support counts are calculated for them.

\begin{enumerate}
    \item \textbf{Unambiguous consensus.} Because all spaced $k$-mer occurrences support the same consensus $k$-mer, the support count is the number of spaced $k$-mer occurrences.
    \item \textbf{Simple ambiguous consensus.} In this case, the produced $k$-mers differ only in a single position. The base support count becomes the number of spaced $k$-mer occurrences that have the matching character at this position. On top of these occurrences, some might have a character at that position which is not used in the produced long $k$-mers. The number of these occurrences is then divided equally between the produced $k$-mers and added to their base support counts.
    \item \textbf{Complex ambiguous consensus.} This case is similar to the previous one. Here the base support count is the number of spaced $k$-mer occurrences that have identical ambiguous position characters as the produced $k$-mer. The leftover occurrences are again split equally between the produced $k$-mers and added to their base support counts.  
    \item \textbf{Undecided consensus.} No consensus $k$-mer is produced, so there is no need to calculate support counts.
\end{enumerate}{}

Figure \ref{consensus-img} shows examples of consensus $k$-mers and their support counts. In cases 1 and 3 there is only one built consensus $k$-mer, so the support count is the number of regular $k$-mers corresponding to the spaced $k$-mer. In case 2, both consensus $k$-mers are supported by two regular $k$-mers, and the last regular $k$-mer is split between them, so the support count for both is 2.5. In case 4 a consensus $k$-mer could not be built, so there is no need to calculate the support count.

\subsection{Memory-efficient and parallel implementation}

For each spaced $k$-mer, \ours\ needs to find its occurrences in the reads. Therefore, before constructing the consensus $k$-mers, all the reads have to be scanned through. Time-wise, it is not efficient to read the input reads separately for all the different spaced $k$-mers. Instead, \ours\ goes through the input reads just once, trying to match the $k$-mers in the reads one by one with all spaced $k$-mers reported by Squeakr. If the read set is reasonably small, \ours\ could store all the reads in memory. It would then be easy to just store the regular $k$-mers as pointers to specific read positions. The stored spaced reads and regular $k$-mer pointers could then be used to find the solid characters and build the consensus $k$-mers. Unfortunately, this would mean that \ours\ worked only with small genomes and small read sets, limiting its usability.

In order to make \ours\ scale to larger genomes, we cannot store the reads in memory nor can we keep all the regular $k$-mers occurring in the reads in memory. To solve this problem, \ours\ keeps a buffer where the regular $k$-mers associated with the spaced $k$-mers are stored, and once the buffer becomes full, the information is stored on disk in temporary files. Then the buffer is emptied, and the scanning of reads for regular $k$-mers continues. The information in the buffer is written to the temporary files in lexicographic order by the spaced $k$-mers. In other words, the regular $k$-mers for the lexicographically smallest spaced $k$-mer come first, and the lexicographically largest last. The size of the buffer $B$ is the number of regular $k$-mers that are kept in the buffer. $B$ is a parameter of \ours\, and can thus be set appropriately by the user depending on the available memory.

The regular $k$-mers are written as a binary file, which makes it easy to use only two bits for every nucleotide character, instead of eight bits if the file was written in a human-readable format. As the usage of two bits enables us to store only four different characters, we are unable to mark unclear nucleotide characters such as the N characters. For this reason, if a gap position in a regular $k$-mer has a character other than A, C, G, or T, one from these four is randomly chosen as a replacement. 

After all the reads are scanned and the regular $k$-mer information has been written to the disk, \ours\ starts building the consensus $k$-mers. Because the information is written to multiple temporary files in lexicographic order,  we can use a technique similar to the multiway merge algorithm. \ours\ starts to read all the files simultaneously, byte by byte. The files are read until all the information (regular $k$-mers) regarding the first spaced $k$-mers of each file is stored in memory. Next, \ours\ determines which spaced $k$-mer among those is lexicographically the smallest, combines the regular $k$-mers for that spaced $k$-mer, and builds the consensus $k$-mers. After that, the information regarding that spaced $k$-mer is discarded, and for every file that contained it, \ours\ continues reading bytes until the information for the next lexicographically smallest spaced $k$-mer in memory. This process continues until no file has more content to be read. The produced consensus $k$-mers are not kept in memory, as they are immediately written to the output file of \ours.

The benefit of this approach is that it is possible to find long $k$-mers even if the input data is too large to keep in memory. As a drawback, this requires more disk space and increases runtime due to the increased disk IO.
At some point, the available disk space may become an issue, which we address by splitting the search and consensus steps into multiple iterations. The spaced $k$-mers are partitioned between these iterations so that each iteration only cares about the $k$-mers specific to it. The temporary files are deleted at the end of the iterations and thus they do not take so much space. The number of iterations affects the runtime because the reads must be read in every iteration to find the iteration specific $k$-mers. Therefore, there is a trade-off between the runtime and available disk space, which can be optimized by the user with a parameter that controls the number of iterations.

\ours\ thus executes sequential iterations with two distinct steps: search step and consensus step. The spaced $k$-mer search with Squeakr is not split into multiple iterations. In the search step \ours\ goes through the input reads and finds occurrences of specific spaced $k$-mers and writes them to temporary files for the consensus step. We have parallelized the search step so that reading the input and associating the regular $k$-mers to spaced $k$-mers is split into multiple threads. The input reads are split into equally sized blocks and each thread is responsible for a single read block.

In the consensus step, \ours\ reads all the temporary files to access the necessary information for consensus $k$-mer building. Our experiments suggested that in this step the bottleneck is the disk IO instead of the actual consensus $k$-mer construction. Because all the temporary files must be read simultaneously due to the multiway merge -like nature of the disk storing implementation, this task cannot be efficiently split between threads.
Thus, we did not utilize parallelism in this step.

\subsection{Time and space complexity}

Here we analyze the time and space complexities of \ours. The whole $k$-mer extraction process can be divided into three steps, and some of them are performed multiple times according to the iteration count. The steps are:
\begin{enumerate}
    \item Squeakr spaced seed search 
    \item \ours\ search step
    \item \ours\ consensus step
\end{enumerate}

\textbf{Squeakr spaced seed search.} Extracting all spaced $k$-mers from a read set with total length $L$ takes $O(L s)$ time where $s$ is the length of the spaced $k$-mer i.e. the number of fixed characters. Inserting a spaced $k$-mer into the counting quotient filter (CQF) takes $O(1)$ time and the contents of the CQF can be enumerated in linear time \cite{cqf} so the total complexity of this step is $O(L s)$. The space complexity of Squeakr spaced seed search is $O(|G|)$ where $G$ is the set of distinct spaced $k$-mers.

\textbf{\ours\ search step.} First, a hashtable is initialized to support membership queries to the set of spaced $k$-mers returned by Squeakr which takes $O(|G|)$ time where $G$ is the extracted spaced $k$-mer set. In each iteration of the search step, \ours\ extracts all the $k$-mers from the reads in $O(L k)$ time.  For each $k$-mer, we check if it corresponds to a stored spaced $k$-mer and if so, it is inserted to the set of $k$-mers for that spaced $k$-mer. This takes $O(\log |G|)$ time per $k$-mer. Therefore the total time complexity of the search step is $O(i L k + L k \log |G|)$ where $i$ is the number of iterations. In this step, the memory contains the hashtable of $|G|$ spaced $k$-mers and the buffer of $B$ regular $k$-mers. Thus the total space complexity is $O(|G|+B)$.

\textbf{\ours\ consensus step.} During the consensus step in total $O(L k)$ regular $k$-mers are read from disk. Let $n$ be the maximum number of regular $k$-mers associated to a spaced $k$-mer. The worst case for consensus $k$-mer generation occurs when there are multiple ambiguous positions. In this case, we compare the regular $k$-mers to each other, which takes $O(n^2k)$ time.  There are at most $\lfloor n/2 \rfloor$ consensus $k$-mers to report and thus the total complexity of this step is $O(L k + gn^2k)$.
The space complexity of this step is $O(n k)$ since at any given time we have the regular $k$-mers associated with the currently processed spaced $k$-mer in memory. 

During an iteration, the temporary files on disk contain the regular $k$-mers associated with the iteration specific spaced $k$-mers. Assuming the regular $k$-mers are evenly distributed over the spaced $k$-mers and the spaced $k$-mers are evenly distributed over the iterations, the maximum total number of $k$-mers stored on disk at any given time is $O(L /i)$.

\section{Experiments and results}
\label{experiments}

We ran experiments on Illumina read set sequenced from the {\em E.~coli} genome with 111x coverage, and a smaller sampled read set with 50x coverage. We also ran experiments on a larger {\em A.~thaliana} genome, with two read sets of coverage 130x and 50x, where the smaller set was sampled from the larger one. The details of the data sets are shown in Table~\ref{data-table}. Additionally, we simulated read sets from the {\em E.~coli} reference genome. Reads were simulated with four different read lengths, 250 bp, 500 bp, 1000 bp, and 2000 bp, and four different substitution error rates, 0.0001, 0.005, 0.010, and 0.020. All combinations of the read lengths and the error rates were used, totaling 16 different data sets. Only substitution errors were simulated, and they were inserted into the reads in a uniformly random manner.
 
We compare \ours\ to DSK \cite{dsk13}, one of the state-of-the-art $k$-mer extraction tools. DSK is a low memory usage program, that supports $k$-mer search for large $k$-mers. This is the reason we chose to compare against DSK as it is also able to extract long $k$-mers. {\em E.~coli} experiments were executed on a machine with 8 cores and 16GB memory, and for the {\em A.~thaliana} experiments we had to use a more powerful machine with 8 cores and 64GB memory. \ours\ was run with the default parameters (Squeakr abundance = 2, solid character minimum count = 2, relative minimum solid character proportion = 0.10), with the exception of using 8 threads, the number of iterations being 32, and having buffer size of 1 000 000 for {\em E.~coli} and 5 000 000 for {\em A.~thaliana}. For \ours\ we used the following hand-picked spaced seed patterns:
\begin{itemize}
    \item \textbf{121-mers:} 6-30-6-15-7-15-6-30-6
    \item \textbf{221-mers:} 6-55-6-40-7-40-6-55-6
    \item \textbf{321-mers:} 6-80-6-65-7-65-6-80-6
    \item \textbf{421-mers:} 6-105-6-90-7-90-6-105-6
    \item \textbf{521-mers:} 6-130-6-115-7-115-6-130-6
\end{itemize}

We evaluated the correctness of the tools by comparing the set of $k$-mers extracted from the reads to the set of $k$-mers occurring in the reference genome which we call the real $k$-mers. We call the intersection of real $k$-mers and $k$-mers extracted from the read set the extracted real $k$-mers. I.e. the extracted real $k$-mers were extracted from the read set and occur in the genome. With these sets of $k$-mers we can compute the precision and recall for the tools:
\[
\textrm{Precision} = \frac{|\textrm{Extracted real $k$-mers}|}{|\textrm{Extracted $k$-mers}|}
\]
\[
\textrm{Recall} = \frac{|\textrm{Extracted real $k$-mers}|}{|\textrm{Real $k$-mers}|}
\]
Finally the F1 score is the harmonic mean of precision and recall:
\[
2\cdot\frac{\textrm{Precision}\cdot \textrm{Recall}}{\textrm{Precision}+\textrm{Recall}}
\]

\begin{table*}[t]

    \caption{Data sets used in the experiments.}
    \label{data-table}

    \centering
    \scalebox{0.95}{
    \begin{tabular}{lllrrrr}
    \hline
    Accession & Organism & Ref. seq. & Genome & Coverage & Number & Avg. read\\
    number & & accession & length & & of reads & length\\
    \hline
    ERR654976 & {\em E.~coli} & GCA\_000005845.2 & 4,641,652 & 50 & 928,274 & 243  \\
    ERR654976 & {\em E.~coli} & GCA\_000005845.2 & 4,641,652 & 111 & 2,120,290 & 243 \\
    SRR5216995 & {\em A.~thaliana} & GCA\_000001735.2 & 119,668,634 & 50 & 20,781,062 & 289 \\
    SRR5216995 & {\em A.~thaliana} & GCA\_000001735.2 & 119,668,634 & 130 & 53,786,130 & 289 \\
     \hline
    \end{tabular}}

\end{table*}

\begin{figure*}[t]
    \centering
    \includegraphics[width=1.0\textwidth]{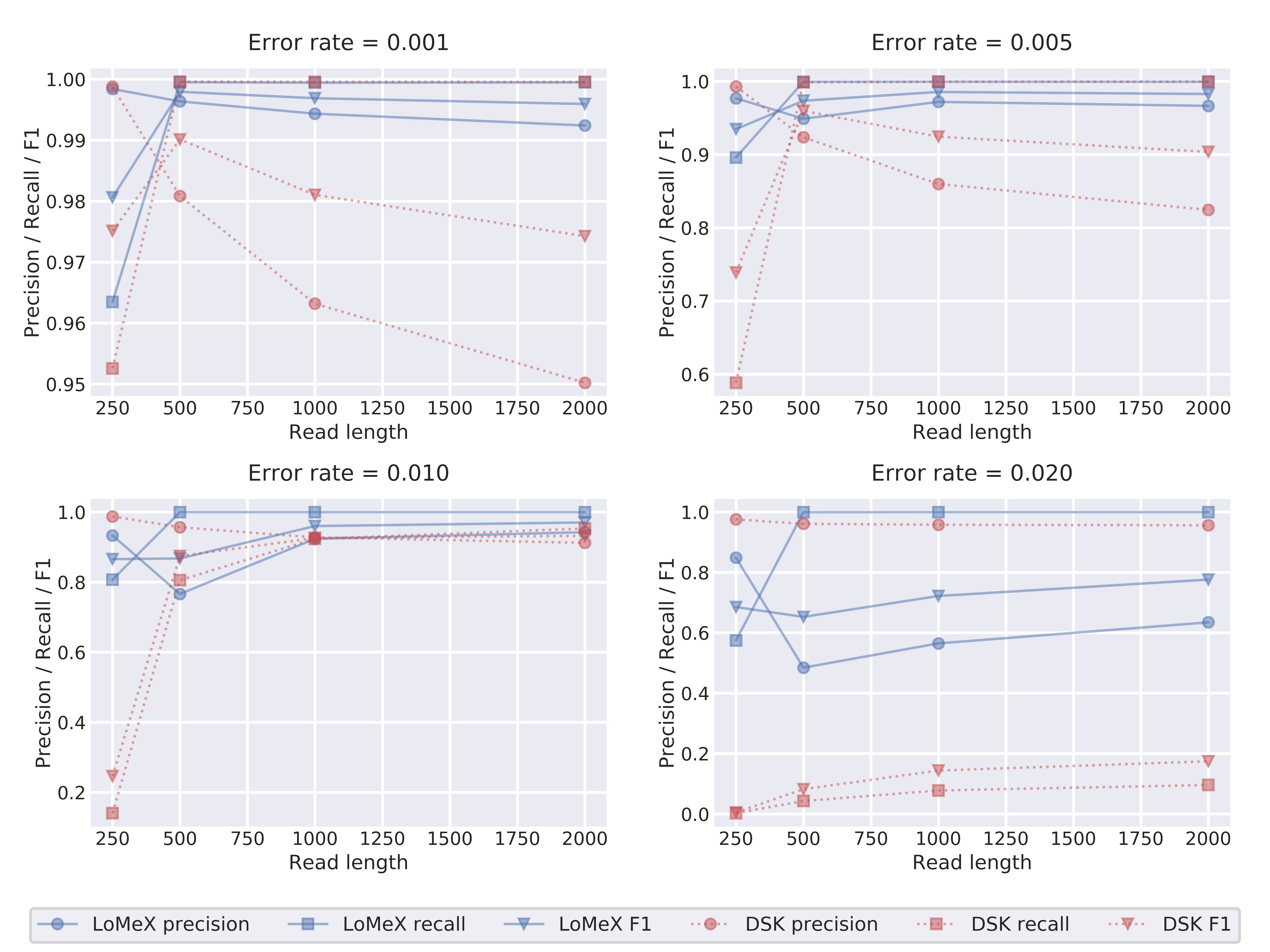}
    \caption{Effect of different error rates and read lengths on LoMeX and DSK 221-mer extraction. This experiment was performed with simulated {\em E. coli} reads.}
    \label{simu-lomex-dsk-comp-error-len-img}
\end{figure*}

We used \ours\ and DSK to extract 221-mers from the simulated data sets to evaluate how well they fared with varying read lengths and error rates. The results of these experiments can be seen in Figure \ref{simu-lomex-dsk-comp-error-len-img}.

\begin{figure*}[t]
    \centering
    \includegraphics[width=1.0\textwidth]{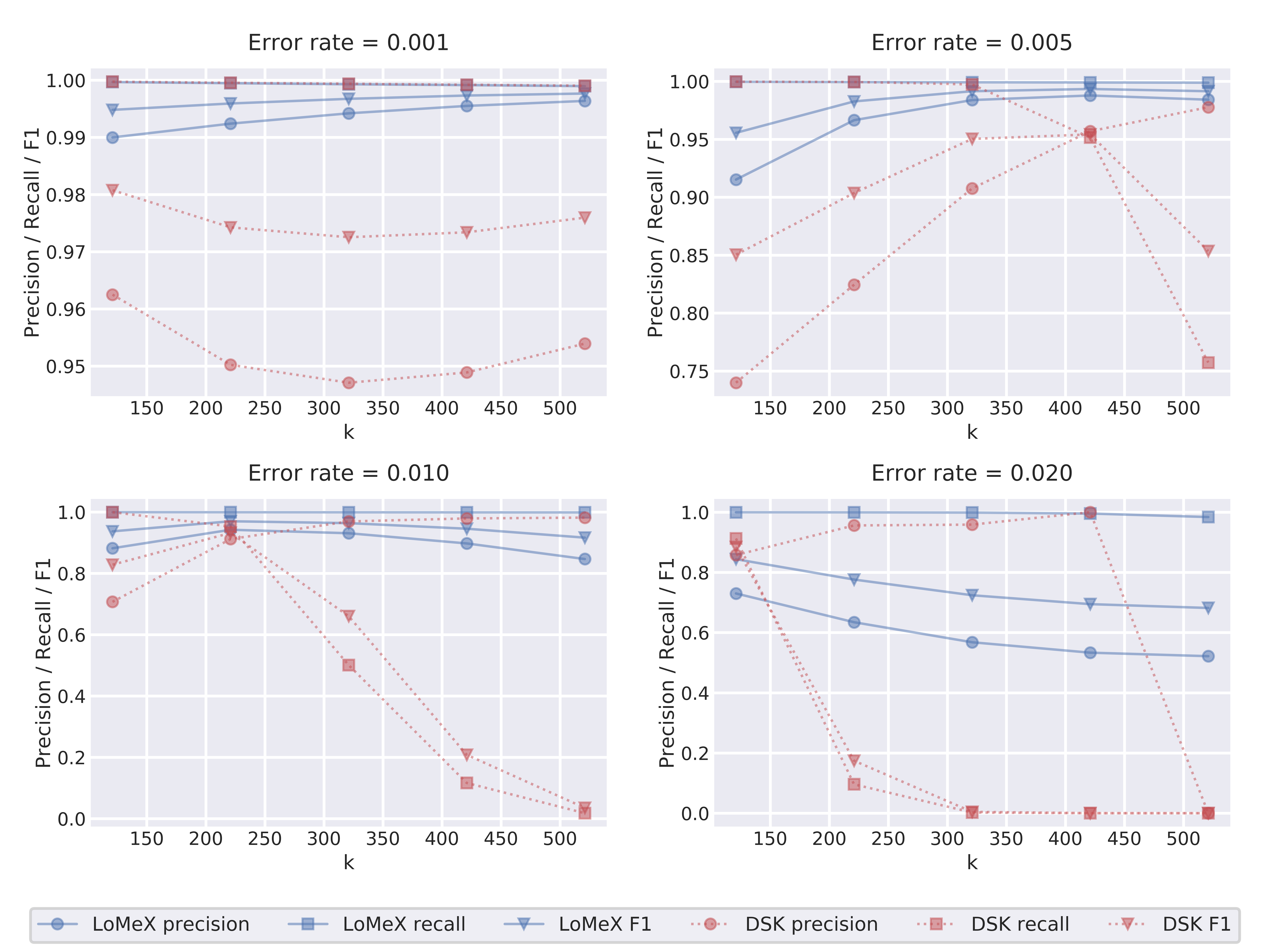}
    \caption{Effect of different error rates and values of $k$ on LoMeX and DSK $k$-mer extraction. This experiment was performed with 2000 bp long simulated {\em E. coli} reads.}
    \label{simu-lomex-dsk-comp-error-k-img}
\end{figure*}

Simulated data sets were also used to evaluate how the value of $k$ would affect the performance of these programs. For this experiment, we used the four read sets with 2000 bp read length. All hand-picked spaced seed patterns were used, and the extracted $k$-mers were compared against DSK extracted $k$-mers. The results of these experiments are found in Figure \ref{simu-lomex-dsk-comp-error-k-img}.

\ours\ was also compared to DSK with real read sets. For these experiments, we used two larger read sets of {\em E.~coli} and {\em A.~thaliana}, and two smaller read sets sampled from the larger ones. In these experiments only 221-mers were extracted. The results can be seen in Table \ref{real-tests-new}, and the runtimes and memory usages in Table \ref{real-tests-new-times}.

\begin{table*}[t]
    
    \caption{
    \ours\ and DSK 221-mer extraction results with different data sets.
    }
    \label{real-tests-new}    
    
    \centering
    \scalebox{0.95}{
    \begin{tabular}{llrrrrrrrr}
    \hline
    Program & Data & Coverage & k & Real     & Extracted & Extracted     & Precision & Recall & F1 \\
            & set  &          &   & $k$-mers & $k$-mers  & real $k$-mers &           &        & score \\
    \hline
    \ours\ & \em E.~coli & 50 & 221 & 4,589,073 & 4,073,788 & 3,971,588 & 0.974913 & 0.865445 & 0.916923 \\
    DSK & \em E.~coli & 50 & 221 & 4,589,073 & 3,948,038 & 3,911,563 & 0.990761 & 0.852365 & 0.916367 \\
    \hline
    \ours\ & \em E.~coli & 111 & 221 & 4,589,073 & 4,798,705 & 4,486,627 & 0.934966 & 0.977676 & 0.955844 \\
    DSK & \em E.~coli & 111 & 221 & 4,589,073 & 4,574,246 & 4,466,413 & 0.976426 & 0.973271 & 0.974846 \\
    \hline
    \ours\ & \em  A.~thaliana & 50 & 221 & 117,866,955 & 127,422,253 & 109,783,497 & 0.861572 & 0.931419 & 0.895135 \\
    DSK\ & \em  A.~thaliana & 50 & 221 & 117,866,955 & 118,800,490 & 105,014,811 & 0.883959 & 0.890961 & 0.887446 \\
    \hline
    \ours\ & \em  A.~thaliana & 130 & 221 & 117,866,955 & 163,363,095 & 115,442,220 & 0.706660 & 0.979428 & 0.820981 \\
    DSK\ & \em  A.~thaliana & 130 & 221 & 117,866,955 & 155,947,309 & 115,468,749 & 0.740434 & 0.979653 & 0.843409 \\
    
    \hline
\end{tabular}}
\end{table*}

\begin{table*}[t]

    \caption{
    Memory usages and runtimes of \ours\ and DSK in 221-mer extraction. These resource usages are related to the results in Table \ref{real-tests-new}. 
    }
    \label{real-tests-new-times}

    \centering
    \scalebox{0.95}{
    \begin{tabular}{llrrrr}
    \hline
    Program & Data set & Coverage & $k$ & Time usage [hh:mm:ss] & Max memory [MB] \\
    \hline
    \ours\ & \em E.~coli & 50 & 221 & 00:11:13 & 8565 \\ 
    DSK & \em E.~coli & 50 & 221 & 00:00:23 & 817 \\
    \hline
    \ours\ & \em E.~coli & 111 & 221 & 00:20:46 & 8847 \\
    DSK & \em E.~coli & 111 & 221 & 00:00:30 & 1649 \\
    \hline
    \ours\ & \em  A.~thaliana & 50 & 221 & 10:35:51 & 12583 \\
    DSK & \em  A.~thaliana & 50 & 221 & 00:28:21 & 8067 \\
    \hline
    \ours\ & \em  A.~thaliana & 130 & 221 & 26:29:17 & 12583 \\
    DSK & \em  A.~thaliana & 130 & 221 & 00:54:30 & 7065 \\
    \hline
\end{tabular}}
\end{table*}

We also experimented with how the choice of the spaced seeds pattern affects the $k$-mer extraction. We chose four random spaced seed patterns of length 221 and compared them to the handpicked 221 length spaced seed. The random spaced seed patterns are the following:

\begin{itemize}
    \item \textbf{Random I:} 1-11-1-2-1-27-1-6-2-3-2-6-1-20-2-6-1-1-1-9-1-4-3-4-1-9-1-1-1-6-2-20-1-6-2-3-2-6-1-27-1-2-1-11-1
    \item \textbf{Random II:} 2-9-1-4-1-8-2-1-1-20-2-1-1-6-1-5-1-23-1-7-1-3-1-8-1-8-1-3-1-7-1-23-1-5-1-6-1-1-2-20-1-1-2-8-1-4-1-9-2
    \item \textbf{Random III:} 1-1-1-6-2-32-1-15-1-6-1-5-1-1-1-2-2-7-1-2-1-10-1-6-1-2-1-2-1-6-1-10-1-2-1-7-2-2-1-1-1-5-1-6-1-15-1-32-2-6-1-1-1
    \item \textbf{Random IV:} 1-10-1-11-1-1-1-2-1-2-1-5-1-9-1-2-1-6-1-7-1-4-1-4-1-19-1-8-1-5-1-5-1-8-1-19-1-4-1-4-1-7-1-6-1-2-1-9-1-5-1-2-1-2-1-1-1-11-1-10-1
\end{itemize}

The results of these experiments are shown in Table~\ref{spaced-seeds-tests-new}. The differences in runtime and memory usage between the spaced seed patterns were negligible.

\begin{table*}[t]
    \caption{
    Comparison between one hand-picked spaced seed pattern and four randomly chosen patterns for 221-mer search. These experiments were done with the 50x coverage {\em E.~coli} reads.}
    \label{spaced-seeds-tests-new}
    
    \centering
    \scalebox{0.95}{
    \begin{tabular}{lrrrrrrr}
    \hline
    Program & $k$ & Real & Extracted & Extracted & Precision & Recall & F1 \\
     &  & $k$-mers & $k$-mers & real $k$-mers &  &  & score \\
    \hline
    Hand-picked & 221 & 4,589,073 & 4,077,016 & 3,972,459 & 0.974355 & 0.865634 & 0.916782 \\
    Random I & 221 & 4,589,073 & 4,109,510 & 3,982,071 & 0.968989 & 0.867729 & 0.915568  \\
    Random II & 221 & 4,589,073 & 4,082,556 & 3,974,210 & 0.973461 & 0.866016 & 0.916601 \\
    Random III & 221 & 4,589,073 & 4,098,478 & 3,978,745 & 0.970786 & 0.867004 & 0.915965 \\
    Random IV & 221 & 4,589,073 & 4,094,384 & 3,977,466 & 0.971444 & 0.866725 & 0.916102 \\
    \hline
    \end{tabular}}

\end{table*}

\ours\ has some parameters that can be tuned to optimize the results of the $k$-mer extraction. Demanding a greater number of occurrences in Squeakr spaced $k$-mer search and higher base count for solid characters will give us fewer reported $k$-mers. This hopefully discards mostly false $k$-mers from the results, leading to higher precision. On the other hand, some weakly supported correct $k$-mers can end up being discarded lowering recall rate. We experimented with differently specified thresholds. In Table \ref{parameter-tests-new} we report the results on how these parameters affect the extracted $k$-mers. Here parameter $S$ is the minimum number of spaced $k$-mer occurrences required by the modified Squeakr. Parameter $N$ is the absolute minimum base count for a base to be considered solid. Again, the differences in runtime and memory usage with the different parameters were negligible so they are not shown.

\begin{table*}[t]

    \caption{Comparison between LoMeX runs for 221-mers with different minimum occurrence and coverage thresholds. The experiment was run using the 50x coverage {\em E.~coli} read set.}
    \label{parameter-tests-new}

    \centering
    \scalebox{0.95}{
    \begin{tabular}{rrrrrrrrr}
    \hline
    $S$ & $N$ & $k$ & Real & Extracted & Extracted & Precision & Recall & F1  \\
        &     &     & $k$-mers & $k$-mers & real $k$-mers &  &  & score \\
    \hline
    2 & 2 & 221 & 4,589,073 & 4,073,790 & 3,971,590 & 0.974913 & 0.865445 & 0.916923  \\
    2 & 3 & 221 & 4,589,073 & 3,389,853 & 3,365,283 & 0.992752 & 0.733325 &  0.843543\\
    3 & 2 & 221 & 4,589,073 & 3,710,129 & 3,615,914 & 0.974606 & 0.787940 & 0.871388\\
    3 & 3 & 221 & 4,589,073 & 3,389,853 & 3,365,283 & 0.992752 & 0.733325 & 0.843543\\
    \hline
    \end{tabular}}

\end{table*}

\section{Discussion}
\label{discussion}

The simulated experiments in Figure \ref{simu-lomex-dsk-comp-error-len-img} give a good overview of how the characteristics of the reads affect \ours. As the read length increases, the input data becomes less fragmented and the total number of $k$-mers in the input data increases, and thus there is more data to construct consensus $k$-mers. For this reason, the recall of \ours\ increases (or at least stays the same), when the read length increases. The same is true for DSK. On the other hand, precision behaves more interestingly. With short reads, it is the highest, but with error rates 0.005 and higher, the precision of \ours\ first drops and then slightly rises as the read length increases. When the total number of $k$-mers in the input reads increases, it is more likely that the same error occurs twice in the reads, and thus precision decreases. However, when the number of $k$-mers in the input data increases further, the minimum character occurrence threshold $c$ in \ours\ starts to increase because \ours\ also requires that a base is present in more than 10\% of regular $k$-mers corresponding to a spaced $k$-mer. DSK precision always declines as the read length increases.

The most promising results were observed in the simulated read experiments. As seen in Figure \ref{simu-lomex-dsk-comp-error-len-img}, the recall of \ours\ is always identical or better than that of DSK, regardless of read length and error rate. On the data sets with the two lowest error rates, \ours\ also has higher precision with reads at least 500 bp long. On the data sets with the higher error rates, \ours\ does not beat DSK in precision. \ours\ F1 score is also higher or nearly identical compared to DSK with all combinations of read lengths and error rates.

Figure \ref{simu-lomex-dsk-comp-error-k-img} shows how well \ours\ and DSK can count $k$-mers of different length with varying error rates and a fixed 2000 bp read length. With the lowest error rate both \ours\ and DSK have similar recalls, but \ours\ has higher precision and F1 score. With 0.005 error rate \ours\ has similar recall for the shorter $k$-mers, but with the longer $k$-mers DSK recall drops noticeably. DSK starts with low precision but begins to catch up with \ours\ as the value of $k$ increases. On data sets with the two highest error rates, DSK has a very low recall with longer $k$-mers. On the other hand, \ours\ is still able to find most of the $k$-mers. With the higher error rates \ours\ starts to lose to DSK in precision. DSK could not find any 521-mers when the error rate was 0.02. 

\ours\ performs very well with simulated data, but it is also important to assess how it performs with real reads. In Table \ref{real-tests-new} we can see 221-mer extraction results of \ours\ and DSK with four different read sets. \ours\ has higher recall with all sets except the 130x one. On the other hand, DSK performed better in precision in all cases. \ours\ is able to extract some harder to find $k$-mers, but as a drawback more false consensus $k$-mers are also built. In Table \ref{real-tests-new-times} we can see the resources used in these experiments. \ours\ is clearly slower than DSK since our program does much more than just search for existing $k$-mers in the read set. 

It is possible that some spaced seeds are better than others for $k$-mer extraction. We did not optimize them, but we checked how an "average" spaced seed would perform by generating four random spaced seeds, and then comparing them to our hand-picked one. The results of this experiment are in Table \ref{spaced-seeds-tests-new}. There is some slight variation in precision and recall, but none of the seeds is significantly better than the others. As future work, it would be interesting to design optimal spaced seed patterns for \ours.
Designing optimal seeds for similarity search has been researched extensively, and various methods have been proposed to solve this problem \cite{seeds05buhler, seeds06kucherov, seeds11speed}. However, these results cannot be directly applied to \ours\ because the studied seed patterns are much shorter than the ones \ours\ requires, and the proportions between fixed and "don't care" characters is much higher when compared to the patterns in \ours. Using more than 31 fixed characters would likely have an effect on the extracted $k$-mers and should also be explored. Extending Squeakr to allow this would be another interesting objective for future work.

The precision of \ours\ is lower than the precision of DSK when using real reads. The precision can be boosted by making the spaced $k$-mer occurrence threshold $S$ and consensus base count threshold $N$ stricter, but this also reduces recall as shown in Table~\ref{parameter-tests-new}. We believe a high recall is ultimately more important because $k$-mers lost by a $k$-mer counter cannot be recovered, whereas erroneous $k$-mers can be detected for example by examining the tip and bubble structure of a de Bruijn graph built on the $k$-mers.

\section{Conclusion}
\label{conclusion}

We have presented \ours, a tool for extracting long $k$-mers from sequencing data. \ours\ uses spaced seeds to successfully find long $k$-mers even in the presence of sequencing errors. Our experiments show that on current real sequencing data, \ours\ has higher recall than state-of-the-art $k$-mer counter, DSK. Furthermore, our experiments on simulated data show that the advantage of \ours\ over DSK increases when the read lengths increase. We expect that the read length of Illumina data will keep increasing and thus \ours\ will become more practical in the future.

Because of the nature of \ours, it is better equipped to handle substitution errors than indel errors. A substitution error only disrupts the spaced $k$-mer matching if the substitution happens at a fixed character position. On the other hand, an indel will disrupt this regardless of where it occurs. Thus \ours\ is not yet ready to process reads from third generation sequencing machines with high rates of indel errors. Nevertheless, the approach pioneered here opens up the possibility to extract long accurate $k$-mers also from these high error rate reads.

\section*{Acknowledgements}

This work was supported by Academy of Finland (grants 308030 and 323233).

\bibliography{bibliography}

\end{document}